\documentstyle[epsf,prl,twocolumn,aps,floats]{revtex}

\begin{document}
\topmargin = -0.6cm
\topmargin = -2.0cm
\overfullrule 0pt

\twocolumn[\hsize\textwidth\columnwidth\hsize\csname
@twocolumnfalse\endcsname
\draft

\title{ 
Direct {\it CP} violation in semi-leptonic and leptonic decays} 

\author{ V. Pleitez}  
\address{
Instituto de F\'\i sica Te\'orica\\
Universidade Estadual Paulista\\
Rua Pamplona 145\\ 
01405-900-- S\~ao Paulo, SP\\Brazil} 
\date{\today}
\maketitle
\vspace{.5cm}

\hfuzz=25pt

\begin{abstract}
We show that direct $CP$ violation in semi-leptonic and leptonic decays
can occur in multi-Higgs doublet extensions of the electroweak standard model
with flavor changing neutral currents.
For pion and lepton decays this $CP$ violating effects cannot be constrained 
by experimental data since up to now the branching ratio of the decays 
$\pi^-$ and $\mu^-$  
have not been measured in laboratory.
\end{abstract}
\vskip2pc]


\newpage
\section{Introduction}
\label{sec:int}

Recently it was pointed out by Kaplan~\cite{kaplan,field} that the comparison
between the polarizations of $\mu^+$ from the decay of $\pi^+$ and of 
$\mu^-$ from the decay of $\pi^-$ could be used in order to verify if $CP$ is 
violated in the $\pi\to\mu\to e$ chain decay. 
Denoting by $A_{\pi^+}$ and $A_{\pi^-}$ the oscillation 
amplitudes for muons from $\pi^+$ and $\pi^-$ respectively it was found from
the $g_\mu-2$ data~\cite{bailey} that~\cite{kaplan} 
\begin{equation}
-0.01<A_{CP}\equiv \frac{A_{\pi^+}-A_{\pi^-}}{A_{\pi^+}+
A_{\pi^-}}<0.02.
\label{acp}
\end{equation}
If this asymmetry is confirmed in the
future, {\it i.e.}, $A_{CP}\not=0$, it means the existence of $CP$ violation 
in pion and/or muon decays. Hence, we can ask ourselves what sort of models 
can produce them. 

The goal of this work is to point out that multi-Higgs doublet extensions of 
the $SU(2)_L\otimes U(1)_Y$ model with flavor changing neutral currents (FCNC) 
in the Yukawa sector and $CP$ violation, through the flavor mixing matrix in 
the interactions with the vector bosons $W^\pm$ and through the scalar 
sector (spontaneous or explicit), imply direct $CP$ violation in 
semi-leptonic and leptonic decays. We also introduce a different way
for counting the physical phases in the
fermion mixing matrices. Although this way coincides with the usual one, it
is more appropriate when there are flavor changing neutral currents in a 
given model.
If there is $CP$ violation but not FCNC the effects are proportional to the
fermion mass and therefore negligible. This of course implies constraints
coming from de neutral meson parameters, notwithstanding, since there
are new mixing angles those constraints do not necessarily imply large
mass for both neutral and charged scalars. For instance masses of the order
of 150 GeV are still possible in models with similar effects to the present 
one~\cite{dumm}.

In the SM~\cite{sm} the only source of $CP$ violation
is the phase in the mixing matrix 
$V_{\rm CKM}$ of the vector charged currents~\cite{km} or, if we enlarge the 
Higgs sector it is possible to implement spontaneous or explicit 
$CP$ violation through the scalar exchange~\cite{tdlsw}. 
As we said before, here we will point out an effect which arises when a model 
has any kind of $CP$ violation and also flavor changing neutral currents 
(FCNC). 

\section{Multi-Higgs extensions of the SM}
\label{sec:one}

In the electroweak standard model (ESM by short) based on the gauge symmetry 
$SU(2)_L\otimes U(1)_Y$ and with only one Higgs doublet, the Yukawa 
interactions in the quark sector are
\begin{equation}
-{\cal L}^q_Y=\bar\psi_L (\Gamma^d\Phi D'_R+ \Gamma^u\tilde\Phi U'_R)+H.c.,
\label{0}
\end{equation}
with $\Phi=(\phi^+,\,\phi^0)^T$, $\tilde{\Phi}=i\tau^2\Phi^*$, and 
$\Gamma^{d,u}$ being
arbitrary complex matrices (Yukawa couplings) in the flavor space, 
$\psi_L=(U',\;D')_L$ denotes the doublet of left-handed fields; $D'_R$ and 
$U'_R$ are gauge singlets; $\tau^2$ is the 
Pauli matrix and primed fields denote symmetry eigenstates.
After the spontaneous symmetry breaking the neutral component of the scalar 
doublet $\phi^0$ is shifted:  $\phi^0=(v+H^0)/\sqrt2$; being $v$ the vacuum 
expectation value (VEV) and $H^0$ a physical scalar field. Then, the Yukawa 
neutral interaction reads in the symmetry basis 
\begin{equation}
-{\cal L}^q_Y=\left( \bar{U'}_LM^uU'_R+\bar{D'}_LM^dD'_R+
H.c.\right)\left(1+\frac{H^0}{v}\right),
\label{1}
\end{equation}
with the quark mass matrices $M^q=v\,\Gamma^q/\sqrt2$. 
Next, we must diagonalize the quark mass 
matrices $M^u,M^d$ by using biunitary transformations
\begin{equation}
V^{u\dagger}_LM^uV^{u}_R=\hat{M}^u,\;\;
V^{d\dagger}_LM^dV^{d}_R=\hat{M}^d,
\label{2}
\end{equation}
with $\hat{M}^u\equiv {\rm Diag}(m_u,\,m_c,\,m_t$) and 
$\hat{M}^d\equiv{\rm Diag} (m_d\,,m_s\,,m_b)$. The physical 
(unprimed fields) states are related to the symmetry eigenstates as follows:
\begin{equation}
U'_L=V^u_LU_L,\; U'_R=V^u_RU_R,\;\;
D'_L=V^d_LD_L,\; D'_R=V^d_RD_R,\;\;
\label{3}
\end{equation}
and with Eqs.~(\ref{2}) and (\ref{3}) the Yukawa interactions in Eq.~(\ref{1}) 
become diagonal in the flavor space
\begin{equation}
-{\cal L}^q_Y=\left( \bar{U}_L\hat{M}^uU_R+\bar{D}_L\hat{M}^dD_R+H.c.\right)
\left(1+\frac{H^0}{v}\right).
\label{4}
\end{equation}
It means that there are no flavor changing neutral currents since 
$\hat{M}^{u,d}$ are diagonal matrices. This also happens in the neutral 
currents coupled to the $Z^0$ gauge boson. In terms of the physical fields the 
lagrangian does not depend at all on the $V^{u,d}_R$ matrices (we will show 
here that this is not the case when we have FCNC) and the matrices
$V^u_L$ and $V^d_L$ appear only as the combination $V_{\rm CKM}=V^{u\dagger}_L
V^{d}_L$ in the charged currents coupled to  $W^+$: 
$\bar{U}_L\gamma^\mu V_{\rm CKM}D_L$ with $U_L=(u,c,t)^T_L$ and 
$D_L=(d,s,b)^T_L$ being mass eigenstates and $V_{\rm CKM}$ being 
an arbitrary unitary matrix.  

Next, it is necessary to determine how many 
phases in $V_{\rm CKM}$ (for simplicity this matrix will be denoted hereafter 
simply by $V$) are measurable. In quantum mechanics only the relative phases 
are important. Therefore, we can redefine the phases of the physical 
left-handed fields~\cite{ceci},
\begin{equation}
\tilde{u}_{\alpha L}= e^{i\varphi(\alpha)}u_{\alpha L},\quad 
\tilde{d}_{\beta L}
=e^{i\varphi(\beta)}d_{\beta L},
\label{5}
\end{equation}
where $\varphi(q)$ are arbitrary real numbers. There are $2N$ of such 
quantities if there are $N$ generations. Under the above transformations we 
have (after absorbing the phases we will forget the ``tilde'' in the fields)
\begin{equation}
\bar{U}_L\gamma^\mu V\,D_L\to \bar{U}_L\gamma^\mu V'D_L =
\bar{U}_L\gamma^\mu F^{u}\, V \, F^{d\dagger}D_L,
\label{6}
\end{equation}
where $F^u\equiv{\rm Diag}(e^{i\varphi(u)},\,e^{i\varphi(c)},\,
e^{i\varphi(t)}, \dots)$ and similarly for $F^d$. In general we can write
$V'_{\alpha \beta}= e^{i[\varphi(\beta)-\varphi(\alpha)]}V_{\alpha \beta}$,
where $\alpha$ and $\beta$ denote an $u$-like and a $d$-like quark, 
respectively. A general $N\times N$ unitary matrix has $N^2$ 
parameters with $N(N-1)/2$ of them taken as Euler angles and the remaining 
ones being phases. 
We see that in the matrix $V'$, $2N-1$ of these phases are 
not measurable. This comes out because we have $2N$ unmeasurable phases 
$\varphi(\beta)$ and $\varphi(\alpha)$ but in $V'$ only the phase differences 
appear and there are $2N-1$ of such quantities (only a common phase 
transformation of all left-handed quarks leaves the elements of $V$ 
invariant). Therefore, $V$ has $N^2-(2N-1)=(N-1)^2$ parameters where 
$N(N-1)/2$ are rotation angles. So, the number of phases in $V'$ is 
$(N-1)(N-2)/2$. 

Although the argument above is correct we will consider a little 
modified one which seems to be more appropriate when the right-handed mixing 
matrices $V^{u,d}_R$ survive in the lagrangian density, like the case in which 
there are flavor changing neutral currents in the theory.
However, it is still necessary to examine how this rephasing affects the 
remaining terms in the lagrangian.

In the ESM the fermion-neutral gauge boson interactions are flavor as well 
as helicity conserving. Thus, there is no effect of the rephasing of the left-
handed fields. The Yukawa interactions, although they are flavor 
conserving, are not helicity conserving. However, it is possible to redefine 
the right-handed quarks exactly with the same phase as the corresponding 
left-handed ones and the Yukawa term remains unchanged too. That is,
\begin{equation}
\tilde{u}_{\alpha R}= e^{i\varphi(\alpha)}u_{\alpha R},\quad 
\tilde{d}_{\beta R}=e^{i\varphi(\beta)}d_{\beta R}.
\label{8}
\end{equation}
In terms of the tilded fields, the lagrangian in Eq.~(\ref{4}) is still 
diagonal, no trace of the phases introduced in Eqs.~(\ref{5}) and (\ref{8}) 
survives. 

As we said before, the Yukawa couplings 
$\Gamma^{u,d}$, or the mass matrices $M^{u,d}$, are
arbitrary complex matrices. It means that they have $2N^2$ real parameters, or 
$N^2$ angles and $N^2$ phases. 
On the other hand, the matrices $V^{u,d}_{L,R}$ are unitary 
matrices that is, each one of them can have up to $N(N+1)/2$ phases. 
The matrices $\hat{M}^{u,d}$ are real and diagonal (with positive 
eigenvalues). It means that the $N^2$ phases of $\Gamma^u$ (or $\Gamma^d$) 
must be absorbed in the $N(N+1)>N^2$ phases of $V^u_L$ {\it plus} the 
phases of $V^u_R$. 
We see that $V^u_L$ and $V^u_R$ do not need to be each one of them general 
unitary matrix, since in this case they have together more phases than the 
number needed to diagonalize $\Gamma^{u}$. 
For instance, if we choose $V^u_L$ to be a general unitary matrix, {\it i.e.}, 
with $N(N+1)/2$ phases, it is sufficient for $V^u_R$ to have only $N(N-1)/2$ 
phases; or vice versa, if $V^u_R$ is the general unitary matrix with 
$N(N+1)/2)$ phases, $V^u_L$ has only $N(N-1)/2$ of them (similarly with the 
$d$-like sector). 

In the context of the ESM or its extensions without FCNC 
both selections are indistinguishable. This can easily be seen 
as follows. In the mixing matrix of the charged currents coupled to the vector 
bosons $W^\pm$ only the pro\-duct $V\equiv V^{u\dagger}_LV^{d}_L$ appears in 
the  lagrangian. 
The matrices $V^{u,d}_R$ do not appear at all in the lagrangian. 
Thus, if we had chosen $V^d_L$ 
($V^u_L$) as the general unitary matrix, independently of the choice of  
$V^u_L$ ($V^d_L$), the matrix $V$ is itself a general unitary matrix with 
$N(N+1)/2$ phases. 
On the other hand, if we had chosen both $V^u_L$ and 
$V^d_L$ as being unitary matrices both with only $N(N-1)/2$ phases, 
the rest of the phases needed to get real and positive mass eigenvalues must 
be  in the matrices $V^{u,d}_R$ and $V$ has only 
$N(N-1)$ phases. The last number has to be equal or less than 
$N(N+1)/2$ which is the maximum number of phases allowed for an unitary 
matrix. Hence, $N(N-1)<N(N+1)/2$ for
$N=2$; but the number of phases in $V$ is again $N(N+1)/2$ for $N\geq3$.
If we use now the phase redefinition of the physical fields in 
Eqs.~(\ref{5}) and (\ref{8}) the observable phases are as usual for $N\geq3$
but for the case of $N=2$ we can have only one phase. It means that we can 
redefine not $2\times(N=2)-1=3$ phase fields but only $2\times(N=2)-2=2$.  
The matrix $V^{u}_R$ has $N(N+1)/2$ or $N(N-1)/2$ phases, if the 
phases of $V^u_L$ are $N(N-1)/2$ or $N(N+1)/2$, respectively, (the 
same for $V^d_R$).
The phases will be observable if the matrices $V^{u,d}_R$ do not disappear 
from the lagrangian as it is the case when the model has FCNC.
Summarizing, for $N\geq3$ we can always choose the number of phases 
equal to $(N-1)(N-2)/2$ in the interaction with the $W^\pm$ gauge boson. 
In the case of $N=2$, if we want not to have phases in $V$ {\it i.e.},
assuming that $V^u_L$ and/or $V^d_L$ are general unitary matrix
it means that $V^u_R$ and/or $V^d_R$ have at least one phase.
Anyway, there will be at least $N(N-1)/2$ for $N\geq 2$ phases in the 
$V^{u,d}_R$ mixing matrices that will be observable if these matrices 
survive in the lagrangian density.

In fact, this way of counting phases in the mixing 
matrices is important in models with additional interactions which are diagonal
in the symmetry basis and have FCNC. For instance if gauge singlets like 
$\bar{\psi}_L\psi_R$ (or vectors like $\bar{\psi}_R\gamma^\mu\psi_R$) are 
allowed. This is the case in the context of the 
$SU(2)_L\otimes U(1)_Y$ model if new generations transform like a vector under 
the gauge symmetry.

With only one Higgs doublet there are no physical charged
fields. However, in extensions of 
the ESM model with several Higgs doublets 
there are physical charged fields. 
If the model has no FCNC the interactions of these fields with the quarks have 
the form
\begin{equation}
\sum_i\left(\bar{U}_LV\hat{M}^d D_R\phi^+_i- \bar{D}_LV^\dagger
\hat{M}^d U_R\phi^-_i\right)+H.c.,
\label{9}
\end{equation}
and we see that the same mixing matrix $V$ of the charged currents coupled
to the vector boson $W^\pm$ appears also in these charged scalar-quark 
interactions. The same $CP$ violating phases appear in both, the Yukawa 
interactions and in the charged currents coupled to the vector bosons.
For two or more doublets the fields $\phi^\pm_i$ are still symmetry 
eigenstates, 
thus it will be possible to have $CP$ violation if the mixing matrix in the 
scalar sector has nontrivial phases, but this is not relevant for the case 
considered here.

In a $n$-Higgs-doublet model with FCNC, the Yukawa term of the lagrangian in 
the quark sector is
\begin{equation}
-{\cal L}_Y=\sum_i\bar\psi_L\left(\Gamma^d_i\Phi_i\right)D'_R+H.c.,
\label{10}
\end{equation}
where $i=1,\cdots n$; plus a similar term in $U'_R$. 
Here $\Gamma^{u,d}_i$ are again arbitrary 
complex matrices in the flavor space. After the spontaneous symmetry breaking 
we have $\phi^0_i=v_i+h^0_i$ and
the $h^0_i$ fields being linear combinations of the physical neutral 
scalars ($h^0_i=\sum_jO_{ij}H^0_j$); the mass matrices are diagonalized as 
follows
\label{11}
\begin{equation}
V^{u\dagger}_L\,\sum_iv_i
\Gamma^u_iV^{u}_R=\hat{M}^u,
\;
V^{d\dagger}_L\sum_iv_i\Gamma^d_iV^{d}_R=\hat{M}^d.
\label{mm}
\end{equation}

The interaction terms with the neutral scalars are of the form 
\begin{equation}
\sum_i\left( \bar{D}_LV^{d\dagger}_L\Gamma^d_{i}V^{d}_RD_R\right)\,h^0_i
+H.c. 
\label{12}
\end{equation}
or
\begin{equation}
\sum_{i,j}\bar{D}_L{\cal O}_{ij}D_RH^0_j+H.c.,
\label{12p}
\end{equation}
where 
\begin{equation}
({\cal O}_{ij})_{\alpha\beta}=\left(V^{d\dagger}_L\Gamma^d_i
V^d_R O_{ij} \right)_{\alpha\beta},\;(i\;\mbox{fixed}).
\label{12pp}
\end{equation}

The matrices $V^d_{L,R}$ diagonalize $\sum_iv_i\Gamma^d_i$ but
not $v_i\Gamma^d_i$  separately for each $i$; 
hence we have flavor changing neutral currents coupled to the neutral scalars. 
Notice that since $\Gamma^d_i$ are arbitrary matrices 
$V^{d\dagger}_L\Gamma^d_iV^{d}_R$ have $N^2$ 
phases in ${\cal N}\leq n-1$ of the $\Gamma^d_i$ matrices. 
We have no more freedom to redefine phases since we have already used 
it in absorbing the phases of the Cabibbo-Kobayashi-Maskawa matrix $V$, as
discussed above.
It means that even in the case of $N=2$ generations we will have four
physical phases in ${\cal N}$ of the matrices $\Gamma^d_i$ appearing in 
the neutral currents via scalar exchange even if the matrix $V^d_R$ has 
only one phase.

The charged Yukawa interactions are of the form
\begin{equation}
\sum_i\bar{U}_LV^{u\dagger}_L\Gamma^d_{i}V^{d}_RD_R \phi^+_i+H.c.,
\label{14}
\end{equation}
and the same number of phases of Eq.~(\ref{12}) survives here too.
Since $\phi^+_i$ are symmetry eigenstate fields we can rewrite 
Eq.~(\ref{14}) in terms of the mass eigenstates $H^+_j$ 
($\phi^+_i=\sum_j{\cal K}_{ij}H^+_j$) 
\begin{equation}
\sum_{i,j}\bar{U}_L  {\cal V}_{ij}D_R H^+_j+H.c.,
\label{15}
\end{equation}
where we have defined 
\begin{equation}
({\cal V}_{ij})_{\alpha\beta}=\left(V^{u\dagger}_L\Gamma^d_{i}
V^{d}_R{\cal K}_{ij}\right)_{\alpha\beta},\;(i\;\mbox{fixed})
\label{16}
\end{equation}
with $\alpha=u,c,t$, $\beta=d,s,b$. 
Notice that the interactions in Eqs.~(\ref{12}) and (\ref{14}) (or 
(\ref{15})) are not proportional to the quark masses; even if 
$O_{ij}$ and ${\cal K}_{ij}$ were complex 
matrices, there are $N^2$ phases in the matrices ${\cal O}$ and 
 ${\cal V}$ in Eqs.~(\ref{12pp}) and (\ref{16}), respectively. 

Concerning the charged leptons, they can be rotated like the 
$d-$like quarks in Eq.~(\ref{3}) but now with $V^l_{L,R}$ instead of 
$V^d_{L,R}$.
In the lepton sector the Yukawa interactions are (with massless neutrinos)
\begin{eqnarray}
-{\cal L}_Y^l&=& \sum_i\left(\bar{\nu}_{L}\,V^{l\dagger}_L\Gamma^l_iV^{l}_R\,
l_{R}\phi^+_i+
\bar{l}_{L}\, V^{l\dagger}_L\Gamma^l_{i}V^{l}_R\, l_{R}\,
h^0_i\right)\nonumber \\ &+&H.c.,
\label{18}
\end{eqnarray}
where we have redefined the neutrino fields so that there is no mixing in the
charged current coupled to the vector bosons $W^\pm$.
The mass matrix for the charged leptons 
$M^l=\sum_i(v_i/\sqrt2)\Gamma^l_i$ is diagonalized as in the 
case of the quarks $V^{l\dagger}_LM^lV^{l}_R=\hat{M}^l$,
with $\hat{M}^l={\rm Diag}(m_e,m_\mu,m_\tau,\cdots)$. 
Hence, the unitary matrices 
$V^l_{L,R}$ diagonalize $M^l$ but not $v_i\Gamma^l_i$ separately.
Although we have redefine the neutrino fields in the charged currents coupled
to the vector bosons $W^\pm$, the same is not 
possible in the interactions with $\phi^\pm_i$. 
Hence, we can see from Eq.~(\ref{18}) that even with massless neutrinos we 
cannot avoid, in general, to have mixing in the charged currents coupled to 
the charged scalars and FCNC mediated by the neutral scalars in the charged 
lepton lepton sector as well. 
If we allow $\Gamma^l_i$ to be general $N\times N$ complex matrices 
we have $N^2$ phases in the Yukawa interactions of the charged 
Higgs in the lepton sector.
 
The currents in Eq.~(\ref{18}) can be written in terms of the physical 
charged scalar:
\begin{mathletters}
\label{18p}
\begin{equation}
\sum_{i,j}\left(\bar{\nu}_{L}\,{\cal V}^l_{ij}\,l_{R}\,H^+_j+
\bar{l}_{L}\,{\cal O}^l_{ij}\,l_{R}\,H^0_j\right)
+H.c.,
\label{18pa}
\end{equation}
with
\label{18pp}
\begin{equation}
({\cal V}^l_{i,j})_{\alpha\beta}=\left(V^{l\dagger}_L\Gamma^l_{i}
V^{l}_R{\cal K}_{ij}\right)_{\alpha\beta}, \;(i\;\mbox{fixed}),
\label{18pb}
\end{equation}
and
\begin{equation}
({\cal O}^l_j)_{\alpha\beta}=\left(V^{l\dagger}_L\Gamma^l_{i}
V^{l}_RO_{ij}\right)_{\alpha\beta}, \;(i\;\mbox{fixed}),
\label{18pc}
\end{equation}
\end{mathletters}
where $\alpha,\beta=e,\mu,\tau$.

\section{Phenomenological consequences}
\label{sec:two}

\begin{figure}[ht]
\centering\leavevmode
\epsfxsize=150pt
\epsfbox{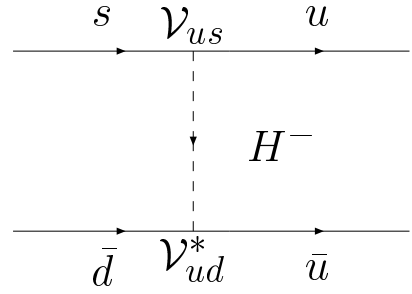}
\vglue -0.009cm
\caption{Charged scalar $H^-$ contribution to 
$K^0_L\to \pi^+\pi^-$.}
\label{fig1}
\end{figure}

An important consequence of this kind of models is that they imply direct $CP$
nonconserving processes. For instance, $\Delta S=1$ processes like
the $K^0_L\to2\pi$ decay. In the ESM only penguin diagrams contribute to this
sort of processes~\cite{wise}. In the present context $CP$ violation arises
because of the interference of the amplitudes of the diagrams shown in
Fig.~\ref{fig1} and that of the standard electroweak model involving $W$ 
bosons. Similar effect exists in hyperon decays~\cite{dono}.

\vglue 0.01cm

More interesting is the case of $CP$ 
violation in semi-leptonic and leptonic decays. For instance, 
$\pi\to l\nu_l$ (particularly when $l=\mu$), 
$\tau\to \mu\nu\bar{\nu}$ and $\mu\to e\nu\bar{\nu}$ 
decays. Usually it is assumed that the $\pi^+$ decay
conserves $CP$. For massless neutrinos the $CP$ mirror image of the
decay $\pi^+\to\mu^+_{LH}+\nu_\mu$ is $\pi^-\to\mu^-_{RH}+\bar{\nu}_\mu$. 
In the first one the helicity of the muon is negative while in the second one 
it is positive. 
Positive pions come to rest then they decay as 
$\pi^+\to\mu^+\nu_\mu$. Next, the muon after traveling some distance comes 
to rest and it decays as $\mu^+\to
e^+\nu_e\bar{\nu}_\mu$. Events of the chain $\pi^-\to\mu^-\to e^-$ are not seen
in this form since negative pions coming to rest in any material are attracted
by a nucleus and captured at a rate too great for the decay be competitive. 
Hence,
it follows that pions decaying in flight in vacuum are required for a $CP$ 
test~\cite{kaplan2}. Similarly for the $\mu^-$ decay.

\begin{figure}[ht]
\centering\leavevmode
\epsfxsize=150pt
\epsfbox{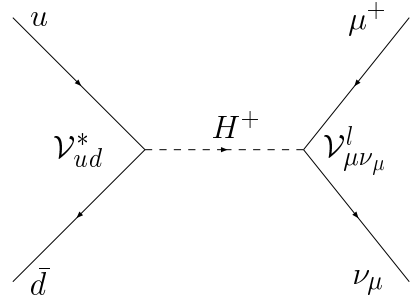}
\vglue -0.009cm
\caption{Charged scalar $H^+$ contribution to 
$\pi^+\to\mu^+\nu_\mu$.}
\label{fig2}
\end{figure}

In models with multi Higgs doublets and FCNC the interference of 
the amplitudes in Fig.~\ref{fig2} and that of the similar diagram 
involving the $W$ boson implies $CP$ violation in $\pi^\pm$ decays. 

\vglue 0.01cm

Here we will use the normalization of Ref.~\cite{pdg} which implies 
$\langle 0\vert A_\mu(0)\vert 
\pi(\vec{q})\rangle = if_\pi q_\mu$ and  
$\langle 0\vert P(0)\vert \pi(\vec{q})\rangle = 
if_\pi m^2_\pi/(m_u+m_d)$~\cite{doli}. We can define the rate asymmetry
\begin{equation}
\Delta_\pi\equiv \frac{\Gamma_{\pi^+}-\Gamma_{\pi^-}}
{\Gamma_{\pi^+}+\Gamma_{\pi^-}}
= \frac{\kappa_\pi}{2\Gamma_{\pi^+}-\kappa_\pi},
\label{22}
\end{equation}
where the difference of the partial width of the respective decays
$\pi^+\to \mu^+\nu_\mu$ and $\pi^-\to \mu^-\bar{\nu}_\mu$ 
($\Gamma_{\pi^\pm}$) is given by the interference term 
$\kappa=4\,{\rm Im}({\cal M}_W{\cal M}_H)$, times 
the phase-space factor (which cancel out in the ratio $\Delta_\pi$);
${\cal M}_W$ (${\cal M}_H$) denotes the invariant amplitude due to the
$W$ vector ($H$ scalar) boson. Thus we have: 
\begin{eqnarray}
\kappa_\pi&\equiv&\Gamma_{\pi^+}-\Gamma_{\pi^-}  
\nonumber \\ && \mbox{} 
=4\,\sum_j
\frac{G_Ff^2_\pi m^3_\pi m_\mu}{8\pi (m_u+m_d)m^2_{H_j}}\;
{\rm Im}(A_j)\sin\Delta\delta,
\label{23}
\end{eqnarray}
where we have defined $\Delta\delta=\delta_+-\delta_-$ with $\delta_+$ 
($\delta_-$) being CP conserving re-scattering phases for the $\pi^+$ 
($\pi^-$), respectively, coming from higher order corrections
to the diagram in Fig.~\ref{fig2}. For instance loop induced correction in
the vertex on the right-vertex in Fig.~\ref{fig2} can arise in the model.
However they must be of the order of $G_Fm^2_\pi\sim10^{-7}$~\cite{sdgv}.
We have also defined 
$A_j=V_{ud}({\cal V}^*_j)_{ud}({\cal V}^l_j)_{\mu\nu_\mu}$.
The contribution of a given scalar $j$ can be suppressed if
${\rm Im}(A_j)\sin\Delta\delta$ is small or;
if the mass $m_{H_j}$ is large. Since the phases $\delta_+$ and $\delta_-$
vanish at leading order we will assume that $\sin\Delta\delta$ is the main 
suppression factor. 
For instance, using $f_\pi\approx0.131$ GeV, $m_u+m_d=10$ MeV,
$m_\mu=106$ MeV~\cite{pdg} 
and for a $j$ fixed $m_{H_j}=100$ GeV we have
\begin{equation}
\frac{\kappa_\pi}{\hbar}\approx
1.4\times 10^{10}\,{\rm Im}(A_j)\,\sin\Delta\delta\;\;\mbox{s}^{-1}. 
\label{23p}
\end{equation}

We do not know what must be the experimental value of $\kappa_\pi$, 
since there is no a direct measure of the difference of the partial width
of $\pi^+$ with respect to $\pi^-$. (It is always measured $\Gamma_{\pi^+}$ 
and it is assumed that the value for $\Gamma_{\pi^-}$ is  the same.)
However if $\kappa_\pi\ll \Gamma_\pi$ we have
\begin{eqnarray}
\Delta_\i&\approx& \frac{\kappa_\pi}{\hbar}\,\frac{\tau_\pi}{2}
\simeq 1.9\times10^{-7}
[1.4\times 10^{10}\,{\rm Im}(A_j)\,\sin\Delta\delta] \nonumber \\
&=&2.66\times10^3 \,{\rm Im}(A_j)\,\sin\Delta\delta
\label{deltanum}
\end{eqnarray}
with ${\rm Im}(A_j)\sin\Delta\delta\approx10^{-10}$ ($j$ fixed), which 
is not an unreasonable value (even if ${\rm Im}(A_j)\lesssim1$) for a 
quantity which arise at higher order, we have
an asymmetry $\Delta_\pi$ of the order of $10^{-7}$ as in Ref.~\cite{sdgv}. 
In fact if we assume $\sin\Delta\delta\sim O(10^{-10})$ there is no contraint
at all on ${\rm Im}(A_j)$ even for a light Higgs scalar ($M_H\sim100$ Gev). 
We stress that $\Gamma$'s matrices in Eqs.~(\ref{18p}), in principle, are 
neither unitary nor hermitian, so the most general constraints come from
perturbation theory: $\vert \Gamma\vert^2/4\pi<1$.
Similar analysis can be done with the $\mu^+$ and $\mu^-$ decays. In
this case we can define in analogy with the $\Delta_\pi$  an asymmetry
$\Delta_\mu$. However, it is not clear for us what is the relation between 
$\Delta_\pi$ and $\Delta_\mu$ and the $A_{CP}$ asymmetry in Eq.~(\ref{acp}).
Notice that the mallness of the CP violation in the $\pi$ and $\mu$ decays
does not implies a small CP violation in the chain $\pi\to\mu\ e$ since
it may exist an CP observable, say $A$, such that 
\begin{equation}
 A(\pi^+\to\mu^+\to e^+)
-A(\pi^-\to\mu^-\to e^-),
\label{ob}
\end{equation}
depends only on the weak phases. In calculating the asymmetry in 
Eq.~(\ref{ob}) muons have to be considered as virtual particles~\cite{2001}.

Contributions to $\epsilon^\prime_K$ at the
tree level constrain ${\cal V}_{us}$; so, 
compatibility with data $Re\,(\epsilon^\prime_K/\epsilon_K)=(28\pm4.1)
\times10^{-4}$ from KTeV~\cite{ktev} and $(18.5\pm7.3)\times10^{-4}$ 
from NA48~\cite{na48,na31} can be obtained by choosing appropriately this
matrix e\-le\-ment (for a $M_H=100$ GeV Higgs scalar) since the 
$\Delta_\pi$ asymmetry does not constrains it too much. 

In the present model there are also contributions to  
$\epsilon^\prime_K$ coming from processes mediated by neutral Higgs bosons 
and in the $K_L\to l\nu_l\pi$ decay 
because of the interference of $\bar{s}\to\bar{u} W^+\to \bar{u}l\nu_l$ and 
$\bar{s}\to\bar{u} H^+\to \bar{u}l\nu_l$. They may be suppressed mainly by
the $CP$ conserving phases. 


\vglue 0.01cm
\begin{figure}[ht]
\centering\leavevmode
\epsfxsize=150pt
\epsfbox{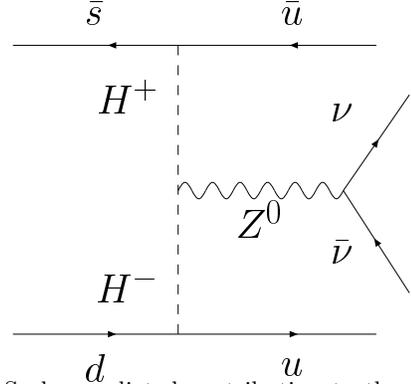}
\vglue -0.009cm
\caption{Scalar mediated contribution to the  
$K_L\to\pi\nu\bar{\nu}$ decay.  }
\label{fig3}
\end{figure}

\section{Rare decays}
\label{sec:thre}

It is worth to make a remark with respect to the rare neutral kaon decays
like $K_L\to\pi^0e^+e^-$~\cite{gaillard} and 
$K_L\to\pi^0\nu\bar{\nu}$~\cite{litt}. Both decays in the standard model 
violate $CP$ in leading order. 
In particular the decay $K_L\to\pi^0\nu\bar{\nu}$ is not only $CP$ violating, 
but also it does not have the potentially 
large $2\gamma$ mediated $CP$-conserving contributions which occur in the
$K_L\to\pi^0e^+e^-$ decay~\cite{sehgal}. 

Denoting the $CP$-violating parameter $\bar{\eta}_{\pi\nu\bar{\nu}}$, it has
been shown that 
$0.1\lesssim \bar{\eta}_{\pi\nu\bar{\nu}}\lesssim1$~\cite{kayser}, 
which is much larger than the corresponding $K\to\pi\pi$ parameters.
Although in the standard model this decay has a branching ratio 
$B(K_L\to\pi^0\nu\bar{\nu})=2.78\times10^{-11}$~\cite{marciano} the 
experimental data give~\cite{pdg}
\begin{equation}
\frac{\Gamma(K_L\to\pi^0 \nu\bar{\nu})}{\Gamma_{\mbox{total}}}
<4.3\times10^{-5}.
\label{fi2}
\end{equation}
It means that this decay can be sensitive to new physics.
In the standard model the main contributions to the decay 
$K_L\to\pi^0\nu\bar{\nu}$ come from penguin and box diagrams. On the other 
hand, in the present model this decay proceeds {\it via}
diagrams like that in Fig.~\ref{fig3}. The interference of 
the diagram in Fig.~\ref{fig3} with a similar one with $H^\pm\to W^\pm$ 
induces $CP$ violating effects.

Independently of the $CP$ issue, using the model independent ratio~\cite{nir}
\begin{equation}
B(K_L\to\pi^0\nu\bar{\nu})<4.4\times B(K^+\to\pi^+\nu\bar{\nu})
\label{oba}
\end{equation}
which is valid even if lepton flavor is not conserved and~\cite{pdg}
$$B(K^+\to \pi^+\nu\bar{\nu})=1.5^{+3.4}_{-1.2}\times10^{-10},$$ 
we obtain
\begin{equation}
B(K_L\to\pi^0\nu\bar{\nu})<6.6\times 10^{-10}.
\label{oba2}
\end{equation}
In the present model the decay $K_L\to\pi^0\nu\bar{\nu}$ arises at the 
tree-level while the decay $K^+\to \pi^+\nu\bar{\nu}$ arises at the
1-loop level. Hence, at first sight it appears that in this model
the inequality in Eq.~(\ref{oba}), which assumes only isospin relations,
can be evaded and $B(K_L\to\pi^0\nu\bar{\nu})>
B(K^+\to\pi^+\nu\bar{\nu})$.
Notwithstanding, notice that the effective interaction Lagrangian arisen from
diagrams like that in Fig.~\ref{fig3} are not
of the four-fermion form $(\bar{s}d)(\bar{\nu}\nu)$ but of a legitimate
six-fermion form. For instance, the strength of the diagram in 
Fig.~\ref{fig3} is proportional to
\begin{equation}
\sum_j\frac{({\cal V}_j)^2_{ud}M_K}{M^4_{H_j}M^2_Z}.
\label{6f1}
\end{equation}

The dimensionless ratio of the strength of the amplitude in Fig.~\ref{fig3}
with respect to the four-fermion 
effective interaction Lagrangian in the standard model, denoted here by 
$A_{SM}$, is (assuming a fixed $j$)
\begin{equation}
R=\frac{M^4_K({\cal V}_j)^2_{ud}/ M^4_{H_j}M^2_Z}{A_{SM}}
\label{ratio1}
\end{equation}
where 
\begin{equation}
A_{SM}=
G_\mu\alpha(M_Z)2\,{\rm Im}\, C/\sqrt{2}\pi\sin^2\theta_W\,\mbox{GeV}^{-2}
\label{sm}
\end{equation}
with the values for the parameters in Eq.~(\ref{sm}) 
given in Ref.~\cite{marciano} we obtain $A_{SM}\approx
3.6\times10^{-11}\,\mbox{GeV}^{-2}$. Hence we have in Eq.~(\ref{ratio1})
\begin{equation}
R\approx 7.7\times 10^4\left(\frac{1\,\mbox{GeV}}{M_{H_j}} \right)^4 
({\cal V}_j)^2_{ud}.
\label{ratio2}
\end{equation}
We see that even a relatively light scalar $M_{H_j}>80$ GeV gives a 
contribution which is $10^{-3}$ smaller than the standard model 1-loop 
contributions.
The decay $K_L\to\pi^0\nu\bar{\nu}$ was considered in two- and three-Higgs 
doublet models with and without FCNC in Ref.~\cite{carlson}. There it was 
shown that the contributions of the charged Higgs bosons for that decay is 
also smaller than the standard model result
and thus unmeasurable. 

In the lepton sector the flavor violation effects via the neutral scalar 
exchange induce not only the usual muonium 
$(M\equiv \mu^+ e^-)$--antimuonium $(\bar{M}\equiv \mu^- e^+)$ 
transition~\cite{hw} but also $CP$ violation, this leaves this system
closer to the neutral kaons~\cite{bp}. Notice that in this model there are 
scalar and pseudoscalar contributions to the $M\to\bar{M}$ 
transition~\cite{muonprd}.  
If the $\vert({\cal V}^{l}_j)_{\mu\nu_\mu}\vert$ matrix element is left 
arbitrary in the pion decay, the $CP$-violation neutral
interactions given in Eqs.~(\ref{18p}) can be large enough to be detected 
by comparing $M\to\bar{M}$ to $\bar{M}\to M$ conversions. 

There are other exotic decays that are induced by this sort of models.
For instance $\mu\to e\gamma$, $\mu\to ee\bar{e}$ and other rare $\tau$ 
decays. The branching ratio of the first decay above is
$B(\mu\to e\gamma)<4.9\times 10^{-11}$~\cite{pdg}. 
In the present model there are
contributions coming from both neutral and charged scalars through the 
interactions in Eqs.~(\ref{18p}). For the charged scalars we 
have~\cite{zee}
\begin{equation}
B(\mu\to e\gamma)=\frac{\alpha}{48\pi}\,\left[\sum_{i,j}
\frac{
({\cal V}^l_j)_{\mu i}({\cal V}^{l\dagger}_j)_{ei}}{M^2_{H_j}G_F}
\right]^2,
\label{mueg1}
\end{equation}
and for accounting the experimental branching ratio we have that 
(for $i,j$ fixed) in Eq.~(\ref{18}) we have
\begin{equation}
\vert({\cal V}^l_j)_{\mu i}({\cal V}^{l\dagger}_j)_{ei} \vert^2
\left(\frac{1\,\mbox{GeV}}{M_{H_j}} \right)^4<1.96\times10^{-16}.
\label{mueg2}
\end{equation}
for a scalar mass of 100 GeV we have that 
$\vert({\cal V}^l_j)_{\mu i}({\cal V}^{l\dagger}_j)_{ei}\vert^2<10^{-8}$. 
When $i=\mu$ this value is compatible with that needed for saturate the 
value of .... The decay $\mu \to ee\bar{e}$ has a 
branching ratio $B(\mu\to ee\bar{e})<1.0\times10^{12}$~\cite{pdg} and it 
is induced in the present model only by neutral (pseudo) scalars. 
It means a constraint only on the 
matrix elements of ${\cal O}^l_j$ and also on the mass of the neutral 
Higgs, but both sort of parameters do not appear in the $\pi\to\mu\nu$ 
decay. This is also valid for the contributions of the neutral scalar 
to the $\mu\to e\gamma$ decay and also to the constraint coming from the
$K_L\to \mu \bar{e}$ decay which has 
$B(K_L\to e^\pm\mu^\mp)<3.3\times10^{-11}$~\cite{pdg}. For the two doublet 
case, the process $K_L\to \mu \bar{e}$ implies scalar masses of the 
30-200 GeV, depending of the ratio of the VEVs~\cite{liu}.
There will be also $CP$ violation in another semileptonic decays
as $B^0\to X\nu_ll$; and also in $p\bar{p}\to l^\pm\nu X$ because of the
interference of $p\bar{p}\to W^\pm X\to l^\pm\nu X$ with 
$p\bar{p}\to H^\pm X\to l^\pm\nu X$ but we will not consider them here
since the exotic leptonic decay seems to be more restrictive. 

There is another source of suppression that we would like to pointed 
out~\cite{fcnit}.
Suppose the case of two doublets. In this case we have two massive neutral
scalars. The charged lepton Yukawa interaction in Eq.(\ref{18}) can be written
as
\begin{equation}
V^{l\dagger}_L\Gamma^l_1V^l_R=\frac{\sqrt2}{v_1}\,\hat{M}^l
-\frac{v_2}{v_1}V^{l\dagger}_L\Gamma^l_2V^l_R.
\label{2dm}
\end{equation}
It means that the vertex in Eq.~(\ref{18pa}) are proportional to
\begin{equation}
V^{l\dagger}_L\Gamma^l_2V^l_R\left( -\frac{v_2}{v_1}O_{1j}+O_{2j}\right)
+\frac{\sqrt2}{v_1}\hat{M}^lO_{1j}.
\label{supre}
\end{equation}
This implies that there are invariant amplitudes that are proportional to
\begin{equation}
\sum_j \left( -\frac{v_2}{v_1}O_{1j}+O_{2j}\right)^2\frac{1}{M^2_{H_j}}.
\label{supre2}
\end{equation}
 
For non-diagonal transitions $\mu\to e,\,\tau\to \mu$ (the later one
appears in the process like $\tau\to\mu e\bar{e}$, which also occurs in these 
sort of models), the mass terms in Eq.~(\ref{supre}) do not contribute. 
Even for diagonal processes, if $v_1$
is of the order of the Fermi scale the mass terms are negligible. It means that
one of the matrix elements can be chosen such that the term between 
parentheses is small, for instance, for the lightest scalar 
$ -(v_2/v_1)O_{1j}+O_{2j}\ll1$ ($j$ fixed). The other
matrix elements are determined by the orthogonality condition but these 
contributions are suppressed by the mass $M_{H_j}$. For more than 
two doublets there will be always some vertices that can be suppressed in 
this way; the other ones can be suppressed by the masses of the scalars.
A similar analysis is valid for the pseudoscalar sector.
Notice that a light neutral scalar contribution to the $K^0-\bar{K}^0$
mass difference must be suppressed by the mixing angles of the right-handed 
matrix $V^d_R$ as it appears in Eq.~(\ref{12}); or/and because of the fact 
that the phenomenological scalar that couples to quarks is different from that
that couples to the charged leptons. 

Finally, let us consider the branching ratio
\begin{equation}
R\equiv\frac{\pi^+\to e^+\nu_e}{ \pi^+\to \mu^+ \nu_\mu}
\label{r}
\end{equation}
which is rather experimentally suppressed and for 
this reason is an important process to test the $\mu e$ 
universality~\cite{emu}. We have that~\cite{pdg} 
\begin{equation}
R^{\rm exp}=(1.230\pm0.004)\times 10^{-4},
\label{emuexp}
\end{equation}
and in the present model we have~\cite{doli}
\begin{equation}
R=R_0\left[\frac{ 1+({\cal V}^*_j)_{ud}({\cal V}^l_j)_{e \nu_e}m^2_\pi/
2^{-\frac{1}{4}}G^{\frac{1}{2}}_Fm_eM^2_H}
{1+({\cal V}^*_j)_{ud}({\cal V}^l_j)_{\mu \nu_\mu}m^2_\pi/2^{-\frac{1}{4}} 
G^{\frac{1}{2}}_Fm_\mu M^2_H}\right]^2,
\label{rteo}
\end{equation}
where $R_0$ is the standard model contribution~\cite{doli}. This
implies that even if $\vert ({\cal V}^*_j)_{ed}({\cal V}^l_j)_{e\nu_e}
\vert\approx1$ a Higgs with $M_H=1000$ GeV, as considered in Eq.~(\ref{23}), 
produces a 0.4\% shift in $R$ for any value of 
$({\cal V}^*_j)_{ud}({\cal V}^l_j)_{\mu \nu_\mu}$. 
However, since neutrinos are not detected in experiments it 
implies similar bound on non-diagonal matrix elements. In models where the
Higgs scalars couple in proportion to the fermions mass the pion decay implies
$M_H=80$ GeV~\cite{ml}; if the couplings are proportional to the mass of a 
heavy fermion we have $M_H>0.5$ TeV~\cite{shanker}. 

\section{conclusions}
\label{sec:con}

We would like to stress that the features we have shown in this work 
can be implemented in other models with complicated Higgs sector and 
intermediate mass scales. An interesting possibility arises when, by
imposing an appropriate discrete symmetry, the scalars
coupled to the leptons are different from the scalars coupled to the quarks.
In this case we have the so called ``leptophilic'' Higgs scalars
since the VEV of the neutral scalars coupled to the leptons may not 
be necessarily of the same order of magnitude than the VEVs which give 
mass to the quarks and vector bosons~\cite{mu1}.

\acknowledgments 
This work was supported by Funda\c{c}\~ao de Amparo \`a Pesquisa
do Estado de S\~ao Paulo (FAPESP), Conselho Nacional de 
Ci\^encia e Tecnologia (CNPq) and by Programa de Apoio a
N\'ucleos de Excel\^encia (PRONEX). I am grateful to D. M. Kaplan and 
C. O. Escobar for useful discussions and to G. Valencia for calling my 
attention to Ref.~\cite{sdgv}.


\begin{references}
\bibitem{kaplan} D. M. Kaplan, Phys. Rev. D {\bf57}, R3827 (1998).
\bibitem{field} The early reference for the use of free leptons to test 
discrete symmetries is J. H. Field, E. Picasso and F. Combley, Sov. Phys. Usp.
{\bf22}(22), 199 (1979).
\bibitem{bailey} J. Bailey {\it et al.}, Nucl. Phys. {\bf B150}, 1 (1979).
\bibitem{dumm} D. G. Dumm, Int. J. Mod. Phys. {\bf A11}, 887 (1996).
\bibitem{sm} S. L. Glashow, Nucl. Phys. {\bf22}, 579 (1967); S. Weinberg, 
Phys. Rev. Lett. {\bf19}, 1264 (1967); A. Salam, {\it in} Elementary
Particle Theory, Ed. by N. Svartholm (Almqviat and Woksell, 1968);
S. L. Glashow, J. Iliopoulos and L. Maiani, Phys. Rev. D{\bf D2}, 1285(1970).
\bibitem{km} M. \ Kobayashi and T. \ Maskawa, Prog. Theor. Phys. {\bf49}, 652 
(1974). 
\bibitem{tdlsw} T.~D. \ Lee, Phys. Rev. D {\bf8}, 1226 (1973); Phys. Rep.
{\bf9}, 143 (1974); S. \ Weinberg, Phys. Rev. Lett. {\bf37}, 675 (1976).
Phys. J {\bf C 11}, 293 (1999) and references therein. 
\bibitem{ceci} For more details see C. \ Jarlskog, in $CP$ {\sl Violation}, 
Edited by C. Jarlskog (World Scientific, Singapore 1989); p.3.
\bibitem{wise} F. Gilman and M. Wise, Phys. Lett. {\bf B93B}, 189 (1980).
\bibitem{dono} J. F. Donoghue, X. G. He and S. Pakvasa, Phys. Rev. D {\bf34}, 
833 (1986).
\bibitem{kaplan2} D. M. Kaplan, presented at the {\sl Workshop on CP 
Violation}, 3--8 July 1998, Adelaide, Australia, hep-ph/9809400.
\bibitem{pdg} D. E. Groom {\it et al.}, The European Physical Journal, 
{\bf C15}, 1 (2000).
\bibitem{doli} J. F. Donoghue and L.-F. Li, Phys. Rev. D {\bf19}, 945 (1979).
\bibitem{sdgv} S. Dawson and G. Valencia, Phys. Rev. D {\bf52}, 2717 (1995).
\bibitem{2001} J. C. Montero and V. Pleitez, in preparation.
\bibitem{ktev} See 
{\footnotesize 
\verb+ http://fnphyx-www.fnal.gov/experiments/ktev/ktev.html+}.      
\bibitem{na48} See also \verb+http://www.cern.ch/NA48/FirstResult+ for the
recent result of the NA48 experiment.
\bibitem{na31} For previous direct CP violation measured in 
$K\to\pi\pi$ decays see also G. D. Barr {\it et al.} (NA31 Collaboration), 
Phys. Lett. {\bf B317}, 233 (1993). 

\bibitem{gaillard} M. K. Gaillard and B. W. Lee, Phys. Rev. D {\bf10}, 867 
(1974); C. O. Dib, J. F. Donoghue, B. R. Holstein and G. Valencia,
Phys. Rev. D {\bf35}, 2769(1987).
\bibitem{litt} L. S. Littenberg, Phys. Rev. D {\bf39}, 3322 (1989);
A. Buras, Phys. Lett. {\bf B333}, 476 (1994).
\bibitem{sehgal} L. M. Sehgal, Phys. Rev. D {\bf38}, 808 (1988).
\bibitem{kayser} B. Kayser, to appear in in the Proceedings of the Summer 
School in High Energy Physics and Cosmology, ICTP-Trieste, June-July, 1995;
hep-ph/9702264 and references therein.  
\bibitem{marciano} W. J. Marciano and Z. Parsa, Phys. Rev. D {\bf53}, R1 
(1996).
\bibitem{nir} Y. Grossman and Y. Nir, Phys. Lett. {\bf B398}, 163 (1997).
\bibitem{carlson} C. E. Carlson, G. D. Dorata and M. Sher, Phys. Rev. 
D {\bf54}, 4393 (1996). 
\bibitem{hw} W-S. Hou and G-G. Wong, Phys. Rev. D {\bf53}, 1537 (1996).
\bibitem{bp} B. Pontecorvo, Sov. Phys. JETP, {\bf6}, 429 (1958).
\bibitem{muonprd} V. Pleitez, Phys. Rev. D; 
hep-ph/9905406 and references therein. 
\bibitem{zee} A. Zee, Phys. Lett. {\bf 93B}, 389 (1980).
\bibitem{liu} J. Liu and L. Wolfenstein, Nucl. Phys. {\bf B289}, 1 (1987).
\bibitem{fcnit} M. M. Guzzo {\it et al.}, hep-ph/9908308.

\bibitem{emu} T. Numao, Mod. Phys. Lett. {\bf A7}, 3357 (1992);
D. A. Bryman, Comments Nucl. Part. Phys. {\bf21}(2), 101 (1993) and references
therein.
\bibitem{ml} B. McWilliams and L-F. Li, Nucl. Phys. {\bf B179}, 62 (1981).
\bibitem{shanker} O. Shanker, Nucl. Phys. {\bf B204}, 375 (1982). 
\bibitem{mu1} J. C. Montero, C. A. de S. Pires and V. Pleitez, 
Phys. Rev. D {\bf60}, 075005 (1999); hep-ph/9812306. 

\end{references}
\end{document}